\documentclass{elsart}
\usepackage{amsfonts}
\usepackage{epsfig}
\newcommand{\re}{{\rm e}}
\newcommand{\ri}{{\rm i}}

\begin{document}
\runauthor{E.A.Kolganova, A.K.Motovilov and Y.K.Ho}
\begin{frontmatter}
\title{Complex Scaling of the Faddeev Equations\thanksref{X}\thanksref{XX}}
\author[IAMS,JINR]{E. A. Kolganova},
\author[JINR]{A. K. Motovilov},
\author[IAMS]{Y. K. Ho}
\thanks[X]{\footnotesize Contribution to Proceedings of the International
           Conference ``Modern Trends in Computational Physics'',
           July 2000, Dubna, Russia}
\thanks[XX]{\footnotesize This work was supported by Academia Sinica,
           National Science Council of R.\,O.\,C.,
           and Russian Foundation for Basic Research}

\address[IAMS]{IAMS, Academia Sinica, P.\,O.\,Box
               23-166, Taipei, Taiwan, ROC}
\address[JINR]{JINR, 141980 Dubna, Moscow Region, Russia}

\begin{abstract}
In this work we compare two different approaches to calculation
of the three-body resonances on the basis of Faddeev
differential equations. The first one is the complex scaling
approach. The second method is based on an immediate calculation
of resonances as zeros of the three-body scattering matrix
continued to the physical sheet.
\end{abstract}
\date{November 24, 2000}
\end{frontmatter}

\section{Introduction}
\typeout{SET RUN AUTHOR to \@runauthor}

The complex scaling method~\cite{BaCom,ReSi} invented in early
70-s remains one of the most effective approaches to calculation
of resonances in few-body systems.  This method is applicable to
an $N$-body problem in the case where inter-particle interaction
potentials are analytic functions of coordinates.  The complex
scaling gives a possibility to rotate the continuous spectrum of
an $N$-body Hamiltonian in such a way that certain sectors of
unphysical sheets neighboring the physical one turn into a part
of the physical sheet for the resulting non-selfadjoint
operator. Resonances appear to be complex eigenvalues of this
operator \cite{BaCom,ReSi} while the binding energies stay fixed
during the scaling transformations.  Therefore, when searching
for the resonances within the complex scaling approach one may
apply the methods which are usually employed to locate the
binding energies.   Some reviews of the literature on the
complex scaling and its many applications can be found, in
particular, in~\cite{Ho,Junker,Reinhard}. Here we only mention
that there is a rigorous mathematical proof \cite{Hag} that for
a rather wide class of interaction potentials the resonances
given by the complex scaling method coincide with the ``true
scattering resonances'', i.\,e. the poles of the analytically
continued scattering matrix in the unphysical sheets.

Along with the complex scaling, various different methods
are also used for calculations of the resonances.  Among the methods
developed to calculate directly the scattering-matrix resonances
we, first, mention the approach based on the momentum space
Faddeev integral equations~\cite{Faddeev63,MF} (see, e.\,g.,
Ref.~\cite{Orlov} and references cited therein). In this
approach one numerically solves the equations continued into
an unphysical sheet and, thus, the three-body resonances arise
as the poles of the continued T-matrix.  Another approach to
calculation of the scattering-matrix resonances is based on the
explicit representations \cite{MotTMF,MotMN} for the
analytically continued T- and S-matrices in terms of the
physical sheet. {}From these representations one infers that the
three-body resonances can be found as zeros of certain
truncations of the scattering matrix only taken in the physical
sheet. Such an approach can be employed even in the coordinate
space \cite{MotMN,ourYaF1}.

To the best of our knowledge there are no published works
applying the complex scaling to the Faddeev equations. Therefore, we
consider the present investigation as a first attempt undertaken
in this direction.  However, the purpose of our work is rather
two-fold. On the one hand, we make the complex scaling of the
Faddeev differential equations.  On the other hand we compare
the complex scaling method with the scattering-matrix approach
suggested in \cite{MotMN,ourYaF1}.  We do this making use of
both the approaches to examine resonances in a model system of
three bosons having the nucleon masses and in the three-nucleon
($nnp$) system itself.

\section{Formalism}

First, we recall that, after the scaling transformation, the
three-body Schr\"odinger operator reads as
follows~\cite{BaCom,ReSi,Hag}
\begin{equation}
\label{SchrScal}
H(\vartheta)=-\re^{-2\vartheta}\Delta_X+
\sum_{\alpha=1}^3 V_\alpha(\re^\vartheta x_\alpha)
\end{equation}
where $\vartheta=\ri\theta$ is the scaling parameter with
$\theta\in{\mathbb R}$. By $\Delta_X$ we understand the
six--dimensional Laplacian in $X\equiv({\bf x}_\alpha,{\bf
y}_\alpha)$ where ${\bf x}_\alpha,{\bf y}_\alpha$ are the
standard Jacobi variables, $\alpha=1,2,3$.  Notation $V_\alpha$
is used for the two-body potentials which are assumed to depend on
$x_\alpha=|{\bf x}_\alpha|$ but not on $\widehat{{\bf
x}}_\alpha={\bf x}_\alpha/x_\alpha$.

The corresponding scaled Faddeev equations which we solve read
\begin{eqnarray}
\nonumber
&&{[-\re^{-2\vartheta}\Delta_X+v_\alpha(\re^\vartheta x_\alpha)-z]
\Phi^{(\alpha)}(z;X)+
V_\alpha(\re^\vartheta x_\alpha)\sum_{\beta\neq\alpha}\Phi^{(\beta)}(z;X)
=f_\alpha(X),}\\
\label{FaddInhom}
&&\qquad\qquad\qquad{\alpha=1,2,3.}
\end{eqnarray}
Here $f=(f_1,f_2,f_3)$ is an arbitrary three-component vector
with components $f_\alpha$ belonging to the
three-body Hilbert space $L_2({\mathbb R}^6)$.

The partial-wave version of the equations (\ref{FaddInhom}) for
a system of three identical bosons at the zero total angular
momentum $L=0$ reads
\begin{eqnarray}
\label{FadPartCor}
{\re^{-2\ri\theta}} H_0^{(l)}\Phi_l(z;x,y) -z\,\Phi_l(z;x,y) +
V({\re^{\ri\theta}}x)\Psi_l(z;x,y)={{f^{(l)}(x,y)}}
\end{eqnarray}
where $x>0$, $y>0$ and $H_0^{(l)}$ denotes the partial-wave
kinetic energy operator,
$$
H_0^{(l)}= -\displaystyle\frac{\partial^2}{\partial x^2}
            -\displaystyle\frac{\partial^2}{\partial y^2}
            +l(l+1)\left(\displaystyle\frac{1}{x^2}
            +\displaystyle\frac{1}{y^2}\right)\,,\qquad l=0,2,4,\ldots,
$$
while $\Psi_l$ stands for the partial-wave component of the
total wave function,
\begin{equation}
\label{FTconn}
         \Psi_l(z;x,y)=\Phi_l(z;x,y) + \sum_{l'}\int_{-1}^{+1}
         d\eta\,h_{l l'}(x,y,\eta)\,\Phi_{l'}(z;x',y')\,.
\end{equation}
Here,
$x'=\sqrt{\frac{1}{4}\,x^2+\frac{3}{4}\,y^2-\frac{\sqrt{3}}{2}\,xy\eta}$
and
$y'=\sqrt{\frac{3}{4}\,x^2+\frac{1}{4}\,y^2+\frac{\sqrt{3}}{2}\,xy\eta}$.
Explicit expression for the geometric function $h_{l l'}(x,y,\eta)$
can be found, e.\,g., in~\cite{MF}.

The partial-wave equations~(\ref{FadPartCor}) are supplied
with the boundary conditions
\begin{equation}
\label{BC0}
\left.\Phi_{l}(z;x,y)\right|_{x=0}=0 \quad \mbox{and} \quad
\left.\Phi_{l}(z;x,y)\right|_{y=0}=0.
\end{equation}
For compactly supported inhomogeneous terms $f^{(l)}(x,y)$ the
partial-wave Faddeev component $\Phi_l(z;x,y)$ also satisfies the
asymptotic condition
\begin{equation}
\label{HeBS}
        \begin{array}{lll}
  \Phi_l(z;x,y) & = & \delta_{l0}\psi_d({\re^{\ri\theta}}x)\exp({\rm i}
   \sqrt{z-\epsilon_d}\,\,{\re^{\ri\theta}y}) \left[{\rm a}_0(z)+
o\left(y^{-1/2}\right)\right] \\
        &+&
 \displaystyle\frac{\exp({\rm i}
 \sqrt{z}\,\,{\re^{\ri\theta}}\rho)}{\sqrt{\rho}}
 \left[A_l(z;y/x)+o\left(\rho^{-1/2}\right)\right],
\end{array}
\end{equation}
For simplicity it is assumed in this formula that the two-boson
subsystem has only one bound state with the energy
$\epsilon_d$, and $\psi_d(x)$ represents its wave function.
The values of ${\rm a}_0$ and $A_l(y/x)$ are the main
asymptotical coefficients effectively describing the
contributions to $\Phi_l$ from the elastic $(2+1\to2+1)$ and breakup
$(2+1\to1+1+1)$ channels, respectively. Hereafter, by $\sqrt{\zeta}$,
$\zeta\in\mathbb C$, we understand the main (arithmetic) branch
of the function $\zeta^{1/2}$.

In the scaling method a resonance is looked for as the energy
$z$ which produces a pole to the quadratic form
$$
{Q(\theta,z)=\left<\left[H_F(\theta)-z\right]^{-1}
f,f\right>}\,
$$
where $H_F(\theta)$ is the non-selfadjoint operator resulting
from the complex-scaling transformation of the Faddeev operator.
The latter operator is just the operator constituted by the
l.\,h.\,s. parts of Eqs. (\ref{FaddInhom}). The resonance
energies should not, of course, depend on the scaling parameter
$\theta$ and on the choice of the terms $f^{(l)}(x,y)$.

In the scattering-matrix approach we solve the same
partial-wave Faddeev equations (\ref{FadPartCor}) with
the same boundary conditions (\ref{BC0}) and (\ref{HeBS})
but for $\theta=0$ and
$$
f^{(l)}(x,y)=-V(x)\int_{-1}^{+1} d\eta\,h_{l 0}(x,y,\eta)\,
\psi_d(x')\sin(\sqrt{z-\epsilon_d}\,y').
$$
The resonances are looked for as zeroes of the truncated
scattering-matrix (see \cite{ourYaF1} for details)
${\rm s}_0(z)=1+2\ri{\rm a}_0(z)$, where the $(1+1\to 1+1)$ elastic
scattering amplitude ${\rm a}_0(z)$ for complex energies $z$ in the
physical sheet is extracted from the asymptotics (\ref{HeBS}).

For numerical solution of the boundary-value problem
(\ref{FadPartCor}\,--\,\ref{HeBS}) we employ its
finite-difference approximation in the hyperradius-hyperangle
coordinates.  A detail description of the finite-difference
algorithm used can be found in Ref.  \cite{KMS-JPB}.

\section{Results}

In the table we present our results obtained for a
complex-scaling resonance in the model three-body system which
consists of identical bosons having the nucleon mass. To
describe interaction between them we employ a Gauss-type
potential of Ref. \cite{ourYaF1}
$$
    V(x)=V_0 \exp[-\mu_0 x^2] + V_b \exp[-\mu_b (x-x_b)^2]
$$
with $V_0=-55$\,MeV, $\mu_0=0.2$\,fm$^{-2}$, $x_b=5$\,fm,
$\mu_b=0.01$\,fm$^{-2}$ and $V_b=1.5$.
The figures in the table correspond to the roots of the inverse
function $[Q(\theta,z)]^{-1}$ for $L=0$ and $l=0$ only taken
into account.  In the present calculation we
have taken up to 400 knots in both hyperradius and hyperangle
variables while for the cut-off hyperradius we take 40\,fm.  One
observes from the table that the position of the resonance
depends very weakly on the scaling parameter $\theta$ which
confirms a good numerical quality of our results.  We compare
the resonance values of the table to the resonance value $z_{\rm
res} =-5.952-0.403\,{\rm i}$\,MeV obtained for the same
three-boson system with exactly the same potentials but in the
completely different scattering-matrix approach of
Ref. \cite{ourYaF1}. We see that, indeed, both the complex
scaling and the scattering matrix approaches give the same
result.

\begin{center}
\begin{tabular}{|c|c|c|c|}
\hline
$\theta$ & $z_{\rm res}$ (MeV) & $\theta$ & $z_{\rm res}$ (MeV) \\ \hline
0.25 & $-5.9525-0.4034\,{\rm i}$ & 0.50 & $-5.9526-0.4032\,{\rm i}$ \\ \hline
0.30 & $-5.9526-0.4033\,{\rm i}$ & 0.60 & $-5.9526-0.4033\,{\rm i}$ \\ \hline
0.40 & $-5.9526-0.4032\,{\rm i}$ & 0.70 & $-5.9526-0.4034\,{\rm i}$ \\ \hline
\end{tabular}
\end{center}

\begin{figure}
\centering
%
\epsfig{file=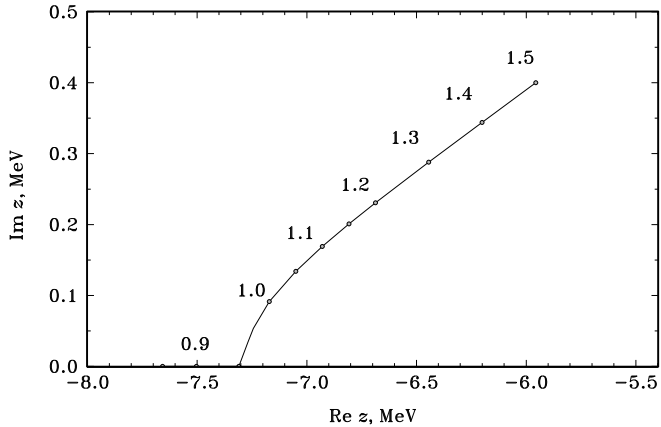,height=11cm}
%
\vspace*{-6.5cm}
\caption{
Trajectory of the resonance $z_{\rm res}$
in the model system of three bosons with
the nucleon masses.  Values of the barrier
$V_b$ in MeV are given near the points marked on the curve.
}
\label{fig-trajectory1}
\end{figure}

We also watched the trajectory of the above resonance when the
barrier amplitude $V_b$ varied (see. Fig. \ref{fig-trajectory1}).
While the complex scaling method was applicable it gave
practically the same positions for the resonance. For the barrier
amplitudes $V_b$ smaller than 1.0 only the scattering-matrix
approach allows to locate the resonance (which finally, for
$V_b<0.85$, turns into a virtual level).

As to the $nnp$ system in the $S$\,--\,state where we employed
the MT\,I--III \cite{MT} potential model, both the methods
applied give no resonances on the two-body unphysical sheet
(see~\cite{ourYaF1}). Moreover, we have found no resonances in
the part of the three-body sheet accessible via the complex
scaling method. Thus, at least in the framework of the
MT\,I--III model we can not confirm the experimental result of
Ref.~\cite{Alexan} in which the point $-1.5\pm
0.3-\ri(0.3\pm0.15)$\,MeV was interpreted as a resonance
corresponding to an exited state of the triton $^3$H.

The triton virtual state can be only calculated within the
scattering-matrix method but not in the scaling approach.  Our
present improved scattering-matrix result for the triton virtual
state is $-2.690$\,MeV (i.\,e.  the virtual level lies 0.47\,MeV
below the two-body threshold).  This result has been obtained
with the MT\,I-III potential on a grid having 1000 knots in both
hyperradial and hyperradial variables and with the value of
cut-off hyperradius equal to 120\,fm.  Notice that some values
for the virtual-state energy obtained by different authors can
be found in~\cite{Orlov} and all of these values are about
0.5\,MeV below the two-body threshold.

\end{document}